# Novel Hybrid DNN Approaches for Speaker Verification in Emotional and Stressful Talking Environments

Ismail Shahin[1,a,*], Ali Bou Nassif[2,b], Nawel Nemmour[1,c], Ashraf Elnagar[3,d], Adi Alhudhaif[4,e], Kemal Polat[5,f]

[1]Department of Electrical Engineering, University of Sharjah, Sharjah, UAE, 27272
[2]Department of Computer Engineering, University of Sharjah, Sharjah, UAE, 27272
[3]Department of Computer Science, University of Sharjah, Sharjah, UAE, 27272
[4]Department of Computer Science, College of Computer Engineering and Sciences in Al-kharj, Prince Sattam bin Abdulaziz University, P.O. Box 151, Al-Kharj 11942, Saudi Arabia
[5]Bolu Abant Izzet Baysal University, Faculty of Engineering, Department of Electrical and Electronics Engineering, Bolu, Turkey

[a]ismail@sharjah.ac.ae, [b]anassif@sharjah.ac.ae, [c]nemours90@hotmail.com, [d]ashraf@sharjah.ac.ae, [e]a.alhudhaif@psau.edu.sa, [f]kpolat@ibu.edu.tr

*Corresponding Author: Ismail Shahin

**ABSTRACT** In this work, we conducted an empirical comparative study of the performance of text-independent speaker verification in emotional and stressful environments. This work combined deep models with shallow architecture, which resulted in novel hybrid classifiers. Four distinct hybrid models were utilized: deep neural network-hidden Markov model (DNN-HMM), deep neural network-Gaussian mixture model (DNN-GMM), Gaussian mixture model-deep neural network (GMM-DNN), and hidden Markov model-deep neural network (HMM-DNN). All models were based on novel implemented architecture. The comparative study used three distinct speech datasets: a private Arabic dataset and two public English databases, namely, Speech Under Simulated and Actual Stress (SUSAS) and Ryerson Audio-Visual Database of Emotional Speech and Song (RAVDESS). The test results of the aforementioned hybrid models demonstrated that the proposed HMM-DNN leveraged the verification performance in emotional and stressful environments. Results also showed that HMM-DNN outperformed all other hybrid models in terms of equal error rate (EER) and area under the curve (AUC) evaluation metrics. The average resulting verification system based on the three datasets yielded EERs of 7.19%, 16.85%, 11.51%, and 11.90% based on HMM-DNN, DNN-HMM, DNN-GMM, and GMM-DNN, respectively. Furthermore, we found that the DNN-GMM model demonstrated the least computational complexity compared to all other hybrid models in both talking environments. Conversely, the HMM-DNN model required the greatest amount of training time. Findings also demonstrated that EER and AUC values depended on the database when comparing average emotional and stressful performances.

## 1. Introduction and Literature Review

As society becomes more connected, human recognition based on biometric systems is gaining a great deal of attention due to the efficient and resilient technological solutions that these systems could offer. Over the past several decades, an increasing number of commercial and noncommercial industries and organizations across the world resorted to these biometric systems to achieve effective natural human–machine communication. Biometric systems have been used in a wide range of applications, ranging from access control to telephone transactions and criminalistics inquiries [1].

Speaker recognition is a voice biometric form of human recognition [2]. Research efforts into this intelligent human–machine interaction are traced back to the 1950s. The field is classified into three major categories: 1) speaker identification (SI), 2) speaker verification (SV), and 3) speaker indexing or diarization (SD) [3]. In fact, speaker identification describes the process of recognizing a speaker from a set of predefined voiceprints, whereas speaker verification aims to evaluate the legitimacy of a speaker against a number of predefined acoustic models. Speaker diarization involves the segregation of a given input speech signal into uniform segments corresponding to the identity of the speaker [3]. On the basis of the type of data used for the training and testing phases, speech can be further classified into two modes: text-dependent and text-independent [4]. Text-dependent modes describe the systems in which the training utterances are identical to those of the testing phrases. Text-independent modes describe the systems where the phrases allocated for the training are not similar to those allotted for testing.

A typical voice communication depends on the linguistic statements as well as on the emotional or stressful facets of the speaker. Verifying a claimed identity in an emotional or a stressful environment by the machine is still challenging for human–machine interfaces particularly because speech is always perceived as a perfect blend of semantic notes allied with emotions besides its paralinguistic characteristics. Subsequently, this paper explores the performance of a text-independent speaker verification system in emotional and stressful acoustic milieus in the light of hybrid DNN models using Mel-frequency cepstral coefficients (MFCCs) and delta–delta features. Verification performance is evaluated among the following hybrid classifiers: HMM-DNN (a novel proposed approach), DNN-HMM, GMM-DNN, and DNN-GMM. The obtained results demonstrate that the proposed HMM-DNN model surpasses all hybrid models in terms of percentage EER and AUC values in all types of emotional and stressful conditions.

Unlike speaker identification, speaker verification is more challenging as it poses a binary classification problem that is based on acceptance or rejection of a claimed identity by an unknown speaker [5]. This requires determining an optimum threshold value at an acceptable level of false rejection and false acceptance. Tuning the decision threshold is a major task in the verification process due to score variability among trials produced by speech content variances, utterance duration, and discrepancies in voice engendered by an emotional state or even acoustical conditions. All of these factors can alter speakers' voice features and can cause difficulties in the verification process.

A large number of studies demonstrate that speaker verification performance in emotional/stressful environments tends to drastically decrease compared to that in neutral-talking environments. This is attributed to two reasons. The first reason relates to the existing problem of discrepancy between training in neutral environments and testing at various emotional/stressful states. This mismatch between the speaker acoustic models and the test utterances causes a bias in verification scores, making the verification process a difficult and challenging task in such environments. The second reason that accounts to the decreased performance of speaker verification in emotional environments might be attributed to the articulating styles of certain emotional states that tend to produce intense intraspeaker vocal variability. According to [6], the verification results of sad or neural emotions surpass those of angry, fear, or happy emotion. By examining the different speech parameters of the aforementioned emotional states, we observe that the pitch ranges corresponding to angry, happy, or fear emotion are largely wider than those of the sadness or neutral states. This suggests that when speakers experience these three sorts of emotions, their articulating styles tend to produce greater intraspeaker vocal variability compared to the vocal variability observed when they are in a sad or neutral state.

Certain studies have focused on speaker verification performance in emotional and stressful acoustic environments using the traditional HMM and GMM classifiers. In his work, Shahin [7] suggested a text-independent speaker verification system on an Emirati database in a neutral-talking environment using the following conventional classifiers: first-order hidden Markov models (HMM1s), second-order hidden Markov models (HMM2s), and third-order hidden Markov models (HMM3s). The experimental results revealed that the HMM3s classifiers demonstrated more reliable verification results compared to those of HMM1 and HMM2 classifiers.

Pillay et al. [8] used GMM-UBM and GMM-support vector machine (SVM) as classifiers in a speaker verification task in a noisy environment. Experiments using white and real-world noises showed that the verification performance based on the use of GMM-UBM and GMM-SVM decreases when the degradation level in the test utterances differs from that in the training sentences. To solve this problem, the authors provided a revised parallel model combination (PMC) technique and proposed a new form of test normalization (T-norm). Results demonstrated that the joint use of GMM-UBM and GMM-SVM drastically improves the verification accuracy in noisy environments.

Shahin in [9] suggested speaker verification architecture that involves merging two recognizers into a single system in neutral and emotional auditory environments. This architecture comprises two cascaded stages: 1) an emotion identifier stage and 2) a speaker verification stage using HMMs in an American language database. Results showed that the speaker verification performance on emotional speech achieved outstandingly better results compared to those of emotion-independent verification tasks.

In [10], Shahin proposed a speaker verification approach in a text-independent setting. The approach entails the emotional features of speakers. The classification is based on two classifiers, namely, HMMs and suprasegmental hidden Markov models (SPHMMs). The proposed architecture involves two cascaded stages that combine the emotional state of speakers and speaker recognizers into a recognizer. Experimental results demonstrated the superiority of the proposed architecture over those reported in previous research works, including emotion-independent and emotion-dependent frameworks that are entirely based on HMMs.

In another study, Shahin and Nassif [11] presented a novel methodology to improve the performance in speaker authentication tasks in neutral and emotional environments. They used three-stage speaker verification architecture, which involved three sequential stages: 1) a gender identification stage, 2) an emotion identification stage, and 3) a speaker verification stage. Model assessment was accomplished on two distinct datasets: the in-house and Emotional Prosody Speech and Transcripts databases. Authors concluded that the newly proposed algorithm enhances the average verification performance compared to the performance of the verification systems that rely solely on gender information or emotion information and of those that lack both gender and emotion cues.

Shahin and Nassif [4] investigated the performance of a speaker verification issue using the third-order circular suprasegmental hidden Markov model (CSPHMM3) as a classifier. The evaluation was performed on an Emirati database using MFCCs as acoustic features. Study findings revealed that CSPHMM3 outperforms the third-order circular hidden Markov model (CHMM3), GMM, SVM, and vector quantization (VQ) for speaker verification tasks in emotional talking environments. The use of CSPHMM3 reduced error rates by 20.6%, 12.8%, and 18.3% based on GMM, SVM, and VQ, respectively.

Over the last decade, deep learning approaches demonstrated a remarkable improvement over the shallow architecture in several domains comprising speech. Investigating DNN and hybrid DNN-based algorithms has become the center of attention of many scholars [12], [13].

Variani et al. [14] proposed a new approach based on DNN as a speaker-intrinsic feature extractor. First, the DNN is trained to classify speakers allocated to the development set at the frame level. Then, the last hidden layer of the DNN is extracted and used for deriving speaker models in the enrollment phase. In the evaluation stage, the decision to admit or reject an identity claim is completed using the cosine similarity between the averaged features, referred to as deep vector or d-vector, of the enrollment utterance with the d-vector of the test utterance.

The authors of [15] presented a method to abstract speaker embeddings from deep learning architecture in a text-independent speaker verification system. The speaker embeddings are computed as a weighted average of speaker-specific acoustic characteristics at the frame level, and their weights are learned using a self-attention technique. Ultimately, the classification of embedding pairs is performed by training a probabilistic linear discriminant analysis (PLDA) classifier. Experimental evaluation on NIST SRE 2016 revealed that the proposed self-attentive embeddings demonstrated better performance than that of the traditional i-vector approach for both long and short utterance tests.

Torfi et al. [16] used a 3D-CNN framework in a text-independent speaker verification task. In both development and enrollment stages, an equal number of utterances per speaker were fed to the DNN to derive speaker models. Authors proposed an adaptive feature learning approach using the 3D-CNN framework for deriving speaker acoustic models. This resulted in seizing the speaker-specific information and constructing a more robust structure that copes with within-speaker variations. The proposed method shows a remarkable performance compared to that of the d-vector approach.

Prasetio et al. [17] proposed a new speaker verification system in emotional stress environments called the emotional variability analysis (EVA)-based i-vector. The speaker model is estimated and generated by following a similar procedure to that of the i-vector technique, but it considers the emotional effect as the channel variability component. The suggested system utilizes the EVA-based i-vector model as the feature extractor and the deep discriminant analysis (DDA) as the channel compensation scheme. The effectiveness of the proposed scheme is assessed by comparing it with the standard i-vector approach using the Speech Under Simulated and Actual Stress (SUSAS) dataset by three diverse scoring methods. The experimental results demonstrated the preeminence of the proposed approach compared to the conventional i-vector system with EER values equal to 4.51%, 4.37%, and 4.08% for cosine similarity scoring (CSS), Euclidean distance scoring (EDS), and Mahalanobis distance scoring (MDS), respectively. The corresponding EER values of the i-vector approach are 6.78%, 6.19%, and 7.63%.

Hourri and Kharroubi [18] proposed a novel technique to use DNN in the field of speaker verification with the aim to enable the DNN to learn the feature distribution in an easier and smoother fashion. In their work, the authors transformed the extracted feature vectors (MFCCs) into enhanced feature vectors, denoted as Deep Speaker Features (DeepSFs). Experimental analysis was performed on THUYG-20 SRE database and results showed that their proposed method outperformed the i-vector/PLDA as well as their baseline system under clean and noisy environments.

In another work, Hourri et al. [19] developed a novel approach, in the context of speaker verification, based on forming two-dimensional CNN filters for the purpose of extracting discriminative speaker information. In addition, the authors proposed novel vectors to recognize speakers called conVectors (i.e. a convolutional neural network vector). Evaluation of their experiments was conducted using the THUYG-20 SRE gender-dependent database in the presence of three noise conditions: clean, 9db, and 0db. Results demonstrated that the use of conv-Vectors yielded the best performance in terms of equal error rates in comparison to the different state-of-the-art approaches for speaker recognition.

Chen et Bao in [20] evaluated their recently suggested neural network called phoneme-unit-specific time-delay neural network (PUSTDNN) which was applied to the x-vector speaker verification system. This neural network model was introduced as an attempt to reduce the impact of variations of speech content which rises the difficulty of the speaker verification task. The proposed approach models each phoneme unit with a distinct time-delay neural network (TDNN). Experiments were carried out on the Fisher, NIST SRE10, and VoxCeleb speech corpora and findings revealed that the PUSTDNN reached the best performance and can attain over 10% relative enhancement in comparison to the x-vector system.

Other studies explored the use of hybrid DNN models in the context of speaker verification. Laskar and Laskar [21] introduced a new approach that integrates the hierarchical multilayer acoustic model (HiLAM) and the DNN-HMM classifier to exalt speaker verification performance in a text-dependent setting. The HiLAM uses the Gaussian mixture model (GMM)-hidden Markov model (HMM) technique and has been exploited for proposing a new data alignment approach for use in DNN training. The authors observed improved system performance and reported an error rate reduction equal to 36.58% on Part 1 of the RSR2015 database.

Bykov et al. [22] employed DNN-HMM in an automated speaker verification system for critical use (ASRSCU) under noisy environments in a text-dependent context. Moreover, their work focused on demonstrating the need to optimize the type and the normalization method of the feature vectors, the number of HMM states and the parameters of Gaussian mixtures, and the technique used to interpret the recognition results. The specificity of the ASRSCU and the DNN-HMM component were taken into account. Detection error tradeoff (DET) curves are utilized to represent the experimental findings, which showed improved speaker recognition efficiency.

Our work differs from other studies in the following points. The authors of [21] and [22] did not examine the emotional or stressful features of audio signals as well as the computational cost of the examined models during the training and testing phases. Moreover, none of the reviewed works inspected the test of statistical significance, such as the Wilcoxon test. They also did not conduct comparison tests between DNN-HMM and other models: HMM-DNN, GMM-DNN, and DNN-GMM. Both studies utilized the DNN-HMM model as the only hybrid classifier inspected. Unlike the previous studies, our work employed EER values, receiver operating characteristic (ROC) curves, and AUC scores rather than EER alone or DET curves alone as performance metrics. Moreover, our work deals with speaker verification in a text-independent context rather than text-dependent. Table I provides a comparative study of the average speaker verification performance demonstrated by the related work in neutral and emotional talking environments. Each reference shows the classifier used, the features extracted, the datasets employed, and the average error rates achieved (EER).

The main contribution of our work is the empirical comparison of speaker verification performances among DNN-HMM, DNN-GMM, GMM-DNN, and HMM-DNN hybrid classifiers. The implementation of HMM-DNN is based on novel architecture. To the best of our knowledge, this research is the first to evaluate and assess speaker verification performance based on the aforementioned hybrid models in emotional and stressful acoustic environments. Furthermore, we conducted pragmatic performance comparisons between these hybrid classifiers and those demonstrated by the following classifiers: HMM, GMM, and DNN. Additionally, we performed six supplementary experiments to validate our findings:

1. The implementation of HMM-DNN is based on novel, newly proposed architecture. The corresponding block diagrams are illustrated in Figs. 5, 6, 7.
2. A systematical evaluation of the verification performance of the hybrid approaches using the stressful SUSAS database [23].
3. Thorough assessments of the verification performances of hybrid models in emotional and stressful acoustic environments.
4. Computational cost assessments of the hybrid models and of the traditional SVM and multilayer perceptron (MLP) classifiers.
5. Conduction and interpretation of the nonparametric Wilcoxon statistical tests between the winning model (HMM-DNN) and all other hybrid models.
6. Comparison of the verification results obtained in the present work with that of previous work.
7. Comparison of the speaker verification performance between emotional and stressful talking environments.

Considering the use of hybrid DNN-based models in a speaker verification task, this work has few common aspects with literature. First, our work used MFCCs as the extracted features just as Laskar and Laskar did in their work [21]. Furthermore, we carried out verification performance comparisons between the DNN-HMM hybrid model and the following shallow classifiers: HMM alone and DNN alone, as solo classifiers, just as Laskar and Laskar did.

The equal error rate (EER) is an evaluation metric usually employed in the verification tasks in order to measure the overall accuracy of a system. EER is the value where the two types of errors: false acceptance rate (FAR) and false rejection rate (FRR) are equivalent. A FRR (type I error) is the percentage of identification samples in which genuine speakers' identity claims are mistakenly rejected, whereas FAR (type II error) refers to the percentage of identification occurrences in which false identity claims are erroneously accepted [9]. In practice, the lower the EER value, the higher the accuracy of the biometric verification model. In general, EER is an iterative algorithm used to decide upon the threshold values for its false rejection and false acceptance rates. The point where both error rates intersect is referred to as an equal error rate. Fig. 1 depicts the graphical representation of the equal error rate in the verification systems - FRR and FAR versus classification threshold [24].

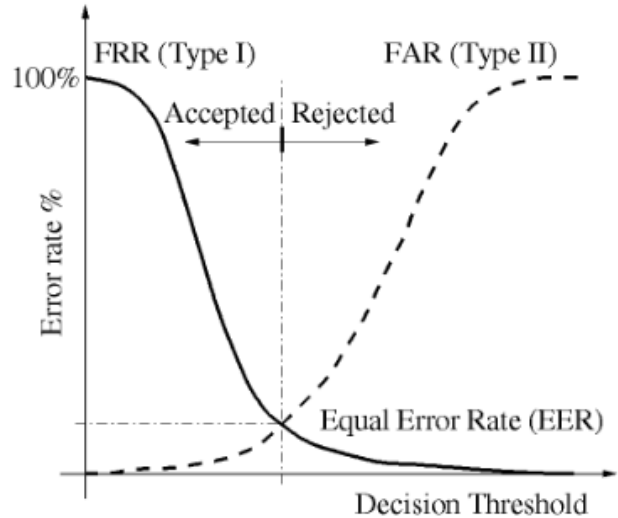

**Figure 1 Graphical representation of equal error rate in verification systems** [24]

The remainder of this paper is organized as follows. Section 2 introduces the emotional speech corpora together with the feature extraction technique. Section 3 provides the background of DNN. In Section 4, the implementation of the hybrid DNN-based models is explained. Section 5 describes the threshold values together with the speaker verification process, whereas Section 6 presents and discusses the experimental results. Ultimately, Section 7 presents the conclusions of this paper.

## 2. Speech Corpora and Feature Extraction

We conducted our experiments using three distinct datasets: A private Arabic Emirati-accented speech dataset (ESD), a SUSAS database, and a Ryerson Audio-Visual Database of Emotional Speech and Song (RAVDESS). Six emotional states were selected: neutral, angry, happy, sad, fear, and disgust. The stressful speaking styles considered were the following: neutral, angry, slow, soft, Lombard, and fast.

### 2.1. Emirati speech database (ESD)

The ESD is a private Arabic Emirati-accented dataset that comprises eight distinct phrases that are widely spoken in the Emirati dialect. Thirty-one nonnative speakers (22 females and 9 males) between 18 and 55 were asked to express each utterance in six different emotions, namely, neutral, angry, happy, sad, disgust, and fear. Each sentence was realized nine times for 2 to 5 s. Table II lists the eight utterances along with their corresponding English translation. This database was recorded at the College of Communication of University of Sharjah, United Arab Emirates. It was obtained with the aid of a speech acquisition board through a 16-bit linear coding analog-to-digital (A/D) converter and sampled at a rate equivalent to 44.6 kHz. The signal samples were pre-emphasized and then segmented into frames of 20 ms, each with an intersection of 31.25% between successive frames.

TABLE I

COMPARATIVE STUDY OF AVERAGE SPEAKER VERIFICATION ACCURACY AMONG THE RELATED WORK

| Reference | Talking Environment | Classifier | Features | Dataset | Average EER |
|---|---|---|---|---|---|
| [9] | Emotional | Two-stage approach based on HMM | Log Frequency Power Coefficients (LFPCs) | Collected speech dataset of American English | 14.41% |
| [10] | Emotional | HMMs, SPHMMs | MFCC | Collected database of American English | 7.75% |
| | | | | Emotional Prosody Speech and Transcripts database (EPST) | 8.16% |
| [7] | Neutral | HMM1s, HMM2s, HMM3s | MFCC | Emirati database | HMM1: 11.5% HMM2: 9.6% HMM3: 4.9% |
| [11] | Emotional | Three-stage framework | MFCC | The collected in-house database | 5.66% |
| | | | | EPST | 6.33% |
| [4] | Emotional | CSPHMM3 | MFCC | Emirati database | 21.75% |
| [14] | Neutral | DNN (d-vector approach) | Log filterbank energy | Collected dataset OK Google phrase | 2.00% |
| [15] | Neutral | DNN | MFCC | NIST speaker recognition evaluations (SRE) SRE16 | Cantonese telephone speech (10s–20s): 5.91% |
| [16] | Neutral | Proposed 3D-CNN | Log-energies | WVU-Multimodal 2013 dataset | 21.1% |
| [17] | Emotional | Proposed EVA-based i-vector model | MFCC | SUSAS database | 4.51% using the cosine distance |
| [21] | Neutral | DNN-HMM | MFCC | Part 1 of the RSR2015 database | Male: 0.20% Female: 0.14% |
| [22] | Noisy | DNN-HMM | Power-Normalized Cepstral Coefficients (PNCC), their energy, and their first and second derivatives | TED-LIUM database | DET curves |

TABLE II
EMIRATI DATASET AND ITS ENGLISH VERSION

| No. | English version | Emirati accent |
|---|---|---|
| 1. | I am leaving now, may God keep you safe | فداعة الرحمن بترخص عنكم الحينه |
| 2. | The one whose hand is in the water is not the same as whose hand is in the fire | اللي ايده في الماي مب نفس اللي ايده في الضو |
| 3. | Where do you want to go today? | وين تبون تسيرون اليوم؟ |
| 4. | The weather is nice, let's sit outdoor | قوموا نيلس في الحوي, الجوغاوي برع |
| 5. | What's in the pot, the spoon gets it out | اللي في الجدر يطلعه الملاس |
| 6. | Welcome millions, and they are not enough | مرحبا ملايين ولا يسدن |
| 7. | Get ready, I will pick you up tomorrow | زهب عمرك طف عليك باجر |
| 8. | Who doesn't know the value of the falcon, will grill it like a chicken | اللي ما يعرف الصقر يشويه |

## 2.2. Speech Under Simulated and Actual Stress (SUSAS)

The SUSAS database is an English public speech database that embraces two subcorpora: a speech under simulated stress and a speech under actual stress [23]. The actual speech under stress entails the speech created while performing either (1) dual-tracking workload computer tasks or (2) subject motion-fear tasks (subjects in roller-coaster rides). The simulated speech involves nine male speakers of three different regional accent classes (Boston accent, General American accent, and New York accent) as well as 11 stressed speaking styles. Both sub-corpora contain 35 different words. Each word is realized two times for every speaker and speaking style. In this work, six stressful speaking styles are considered (neutral, angry, slow, soft, Lombard, and fast). All speech tokens were sampled using a 16-bit A/D converter at a sample rate of 8 kHz.

## 2.3. The Ryerson Audio-Visual Database of Emotional Speech and Song (RAVDESS)

The RAVDESS is a public database expressed in English [25]. It includes 24 professional actors (12 females and 12 males) enunciating two lexically matched phrases in a neutral North American accent: 1) "Kids are talking by the door" and 2) "Dogs are sitting by the door." The RAVDESS encompasses 7,356 files in total, 1,440 out of which are audio speech files (60 trials per actor × 24 actors). Moreover, the database includes seven different emotional sets. This work considers only six types of emotions: neutral, anger, happiness, sadness, fear, and disgust. All 1440 audio files were down-sampled to 16 KHz.

## 2.4. Feature extraction

In this work, the coefficients that characterize the phonemic content of the audio signals are called the MFCCs. These coefficients have been the most widely used acoustic features in speech recognition [26], [27], emotion recognition [28], [29], and speaker recognition [10], [30], [31], [32]. This is due to the high-level estimation of human auditory perception that these coefficients could offer [11], [33]. The dimensional audio features are extracted using libROSA, a Python package for music and audio analysis.

A frequency of a pure tone in Hertz ($f$) is converted into Mel frequency ($m$) using the subsequent equation [34]:

$$m = 2595 \ log_{10} \left(1 + \frac{f}{700}\right). \tag{1}$$

The delta and delta–delta coefficients are extracted and appended to the MFCC feature vector in ESD and SUSAS databases to enhance verification accuracy but not in RAVDESS datasets. Experimental trials demonstrate that the verification accuracy does not improve when the delta and delta–delta features are utilized in RAVDESS datasets. The delta features are computed using the following formula [35]:

$$d_t = \frac{\sum_{n=1}^{N} n(c_{t+n} - c_{t-n})}{2 \sum_{n=1}^{N} n^2}, \quad (2)$$

where $d_t$ is a delta coefficient from frame $t$ computed in terms of the static coefficients $c_{t+N} - c_{t-N}$. The value of $N$ is usually taken as 2. The delta–delta coefficients are computed in the same manner, but they are calculated from the delta rather than from the static coefficients.

## 3. Background of Deep Neural Networks

A DNN is feed-forward artificial neural network architecture that employs more than one hidden layer between its input and output layers to classify a feature or learn a representation [36]. When the problem requires the classification of a feature x among Q categories, the DNN estimates the probabilities $p_j, j \in \{1, ..., Q\}$ over the classes Q. The input features of a DNN are usually a representation of the time–frequency of the input audio signal such as MFCCs and Mel-filter banks. To form the DNN input, feature frames are concatenated in a sliding window. Fig. 2 illustrates the DNN architecture for a classification task. For $H$ layers, DNN computes a nonlinear function, as follows [36]:

$$g_w(x) = g_H\left[W_{HGH}(W_{h-1} \ldots g_1(W_1 x))\right], \quad (3)$$

where $x$ denotes the input features, $g_h, h = 1, ..., H$ represents the activation functions, and $W_h, h = 1, ..., H$ are the DNN weights.

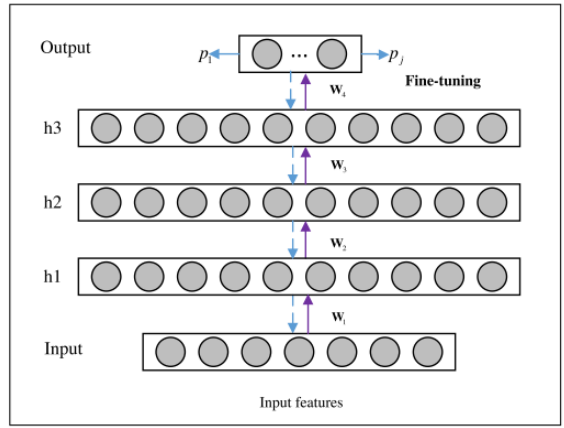

FIGURE 2. DNN architecture for classification [36]

For a classification problem, the output is calculated using the Softmax activation function. The weights of a DNN are trained by minimizing the cross-entropy cost function provided by [36]

$$C = -\sum_{j=1}^{Q} q_j \log p_j, \quad (4)$$

where $C$ corresponds to cross-entropy cost function, $p_j$ denotes the output of the Softmax, and $q_j$ represents the corresponding target.

## 4. Hybrid Model Design

The database in each hybrid model is partitioned into a development (training) set and an evaluation set. The evaluation set consists of enrollment and testing, and it represents approximately 20% of the database as a whole. The following subsections thoroughly describe the implementation of the DNN-based hybrid models using the Emirati database. Twenty-four out of thirty-one speakers are designated (7 males and 17 females) for the training phase, whereas seven (2 males and 5 females) of them are allocated to both enrollment and testing stages.

### 4.1. DNN-HMM Modeling

#### 4.1.1 DNN-HMM Training

First, we generate the GMM-HMM acoustic models for each speaker using the neutral utterances designated for the training phase. Each speaker model uses 5 states and 64 Gaussian mixtures with a diagonal covariance type. Then, the frame-level state labels are derived from the GMM-HMM models by force alignment using the maximum likelihood estimation (MLE) and the Viterbi algorithm [36]. The input data for DNN training correspond to n successive MFCC frames. The HMM states produced by force alignment represent the data labels of the DNN. Next, all training utterances along with their labeled state sequences serve as inputs for training the DNN, which consists of dense layers. The nodes of the Softmax layer (the output layer) are composed of the total numbers of speakers designated for the training multiplied by the number of HMM states (24 speakers × 5 HMM states = 120 output units). Ultimately, the DNN is trained using the backpropagation (BP) algorithm. As shown in Fig. 3, each output neuron of the DNN is trained to estimate the posterior probability of the HMM state based on observations. The cross-entropy training procedure is found in Section 3.

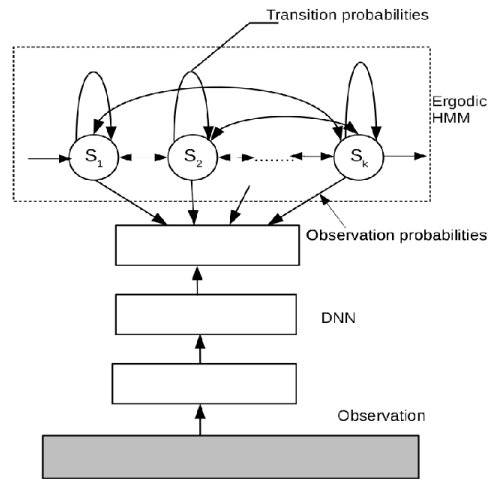

**FIGURE 3. Architecture of the DNN-HMM model** [37]

#### 4.1.2 DNN-HMM Decoding

In the testing stage, the well-trained DNN-HMM model, which is discussed in Section 4.1.1, is loaded, and the posterior probabilities of states given the test observations are produced. Because the HMM requires the use of likelihood rather than the posterior probability during the decoding procedure, we first convert the posterior probability into likelihood using the Bayesian rule, as shown below [38]:

$$p(o_t|q_t) = \frac{p(q_t|o_t)\,p(o_t)}{p(q_t)}. \tag{5}$$

In our study, the prior probability of each HMM state, $p(q_t)$, is computed from the occurrences of the training set, and $p(o_t)$ is allotted a constant value because the observation feature vectors are viewed as independent from each other. Ultimately, the Viterbi algorithm helps in decoding the HMM state sequences, and the speaker verification task is performed. Section 5 explains the verification procedure in detail.

**4.2. DNN-GMM Modeling**

4.2.1 DNN-GMM Training

Initially, we built the GMM acoustic models for each speaker. Each speaker model uses 16 components with a diagonal covariance type and an iteration number set to a maximum of 1000 using the expectation-maximization (EM) iterative algorithm. The number of Gaussians $N_g$ influences each extracted frame feature in a different way. Therefore, if we allot each frame feature to a single Gaussian distribution with maximum likelihood among all Gaussians in the GMM, we can utilize the single Gaussian as a new subclass label to substitute the original speaker label as follows [36]:

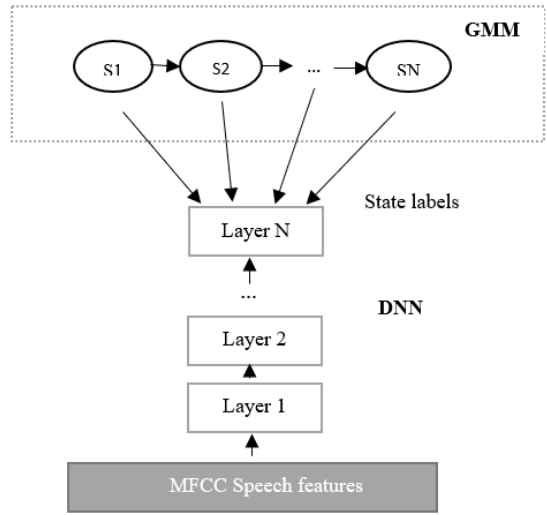

**FIGURE 4. Architecture of the DNN-GMM model**

$$S_{i^*}^j(X) = argmax\, L_i^j(X), \tag{6}$$

where $S_{i^*}^j(X)$ denotes the subclass label of the input training MFCC features $X$, $i^*$ represents the index of Gaussian with maximum likelihood in the GMM model, $j$ is the index of the speaker class, and $L_i^j(X)$ is the likelihood of $i$th Gaussian in $j$th speaker class. The total number of subclasses is $Q \times N_g$, where Q is the number of speakers in the training set (24 speakers × 16 components = 384 subclasses). In this study, the DNN is trained to predict the probabilities over the total $Q \times N_g$ GMM subclasses ($S_1^1,\ldots,S_i^j$, $\ldots,S_{N_g}^Q$) with MFCC + delta–delta coefficients as inputs, resulting in the DNN-GMM hybrid model. Fig. 4 shows the block diagram of the DNN-GMM model.

4.2.2 DNN-GMM Decoding

When an emotional test utterance is fed to the DNN-GMM model trained in Section 4.2.1, posterior probabilities over the GMM subclasses are generated. Finally, EER and AUC scores are produced using the roc_curve built-in function, and speaker verification is achieved (see Section 5).

**4.3. HMM-DNN modeling**

The newly developed block diagrams depicting the development, enrollment, and testing phases of the HMM-DNN and GMM-DNN models are shown in Figs. 5–10.

4.3.1 HMM-DNN Training

At the beginning of the HMM-DNN training phase, every single MFCC vector of the training data competes with all the HMM acoustic models derived for each speaker designated for the development phase. Every speaker $Sp$ pronounces a number of utterances $U_N$. After the feature extraction phase, every MFCC feature of a given speaker $Sp$ is scored against all HMM speaker models in the training (HMM$_{Sp1}$, …, HMM$_{SpN}$) using the "score" Python built-in function. Hence, we acquire an array of vectors for each speaker. Each individual vector contains the log likelihood scores computed and has a length equivalent to the number of speakers allocated to the training stage. The array (containing all the vectors) has a length equivalent to the total number of MFCC features associated with every speaker. At a later stage, each MFCC is competed with its corresponding HMM speaker model. Then, the difference between the recently obtained log-likelihood score and the stored scores for that speaker is calculated. Consequently, a new vector is created. Eventually, we obtain an array of vectors belonging to every single speaker. The newly generated arrays of vectors along with their corresponding speaker labels are used as inputs to train the DNN. Fig. 5 illustrates a graphical representation of the architecture of the implemented HMM-DNN model during the training phase.

4.3.2 HMM-DNN Decoding

During this stage, the enrollment utterances issued from the corresponding speakers are used to retrain the HMM-DNN model (trained in Section 4.3.1). First, we calculate the log likelihood obtained using the "score" function between each enrollment MFCC feature of every speaker against all HMM enrollment speaker models. Subsequently, every speaker (in the enrollment) obtains an array of vectors where each vector contains the log likelihood scores calculated with a length equivalent to the number of speakers designated for the enrollment phase. Next, each MFCC feature of every speaker in the enrollment stage is competed with its relevant HMM speaker model then the difference between this log likelihood and the stored loglikelihood scores of that speaker is calculated. A new vector is obtained. As the number of speakers in the enrollment stage is fewer than that in the training stage, we padded with zeros the vector generated so that it has the same length as the total number of speakers allocated to the training stage. Eventually, each speaker allocated for enrollment has its corresponding array of vectors. These arrays and their corresponding speakers' labels form the input of the pretrained DNN. Afterwards, the pretrained HMM-DNN is loaded, and the last Softmax layer is replaced by a new dense layer as depicted in Fig. 6. This layer has a number of nodes equivalent to the number of speakers dedicated to the enrollment and test stages. Eventually, the resulting DNN is trained using the enrollment data.

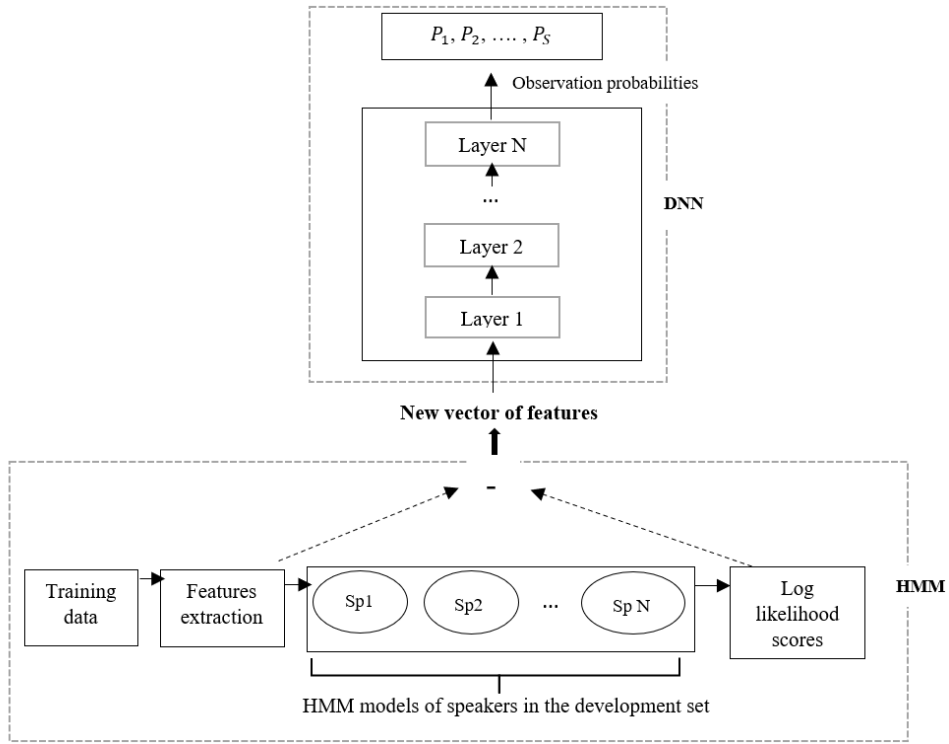

**FIGURE 5.** Architecture of HMM-DNN in the training phase

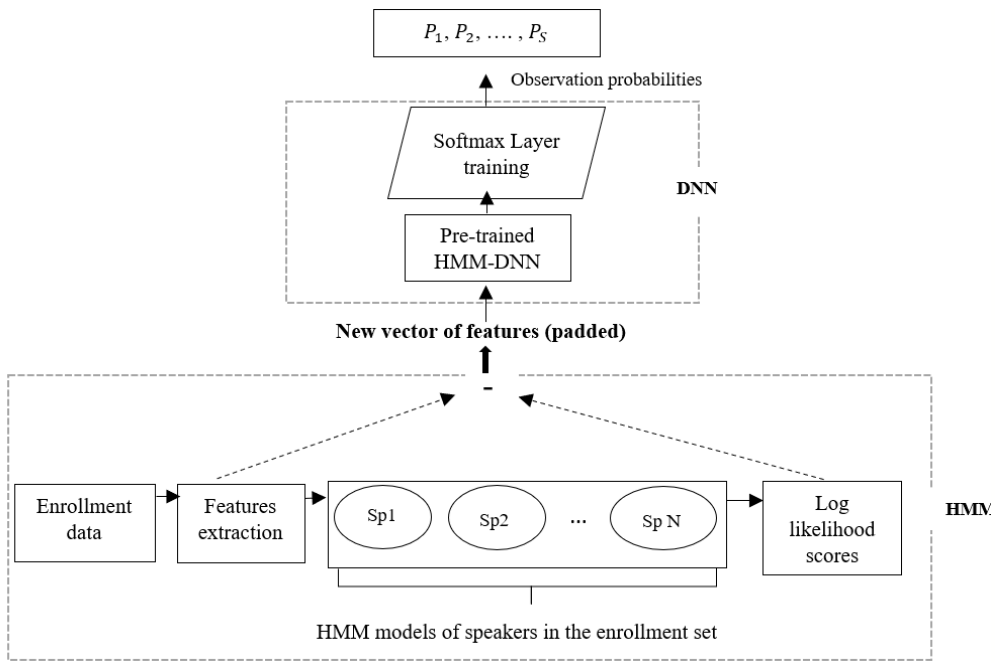

**FIGURE 6.** Architecture of the HMM-DNN model in the enrollment phase

Lastly, in the testing stage, the test samples, issued from the different emotional states, are fed into the pretrained HMM-DNN hybrid model, discussed in Section 4.3.2, and the verification of speakers is performed (see Section 5). Fig. 7 illustrates the architecture of HMM-DNN in the testing phase.

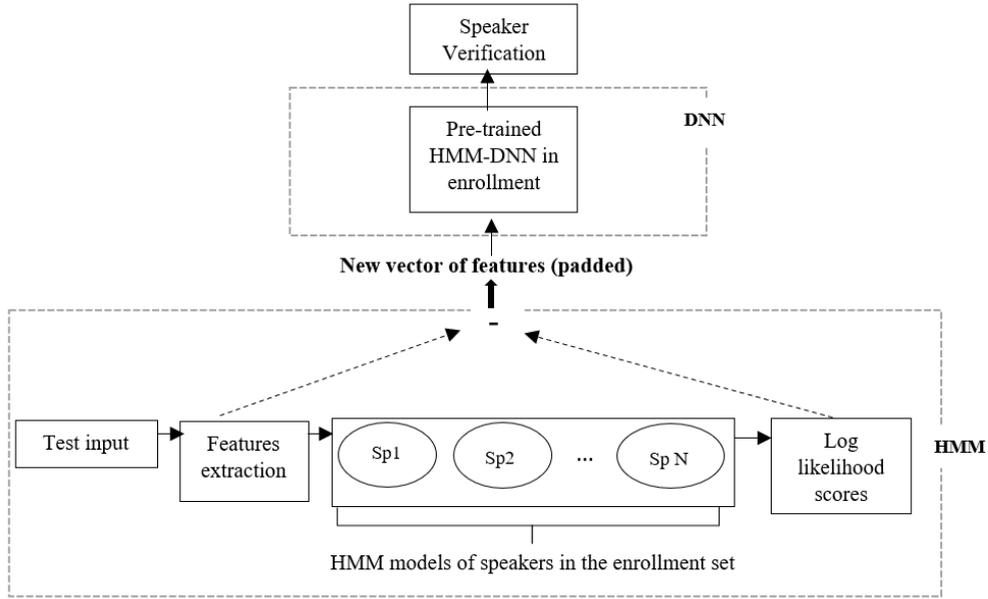

**FIGURE 7. Architecture of the HMM-DNN model in the testing phase**

**4.4. GMM-DNN modeling**

4.4.1 GMM-DNN Training

The same approach employed in Section 4.3 is used for GMM-DNN modeling. Initially, every single MFCC feature frame of the training data competes with the corresponding GMM speaker acoustic model in the development set, obtaining a vector of log likelihood scores. Then, the difference between the MFCC features and the saved log likelihood scores is calculated, and a new vector is produced. The generated vector along with the corresponding speaker names is used as inputs to train the GMM-DNN model. Fig. 8 demonstrates the block diagram of the GMM-DNN hybrid model during the training phase.

4.4.2 GMM-DNN Decoding

During enrollment, the pretrained GMM-DNN in Section 4.4.1 is loaded, and the last Softmax layer is replaced by a new dense layer, as shown in Fig. 9. This layer has the same number of nodes as the number of speakers in the enrollment set. Next, the resulting GMM-DNN model is trained using the enrollment data. During the testing phase, the emotional test samples are fed to the pretrained GMM-DNN, and speaker verification is performed. Fig. 10 shows the architecture of GMM-DNN in the testing phase.

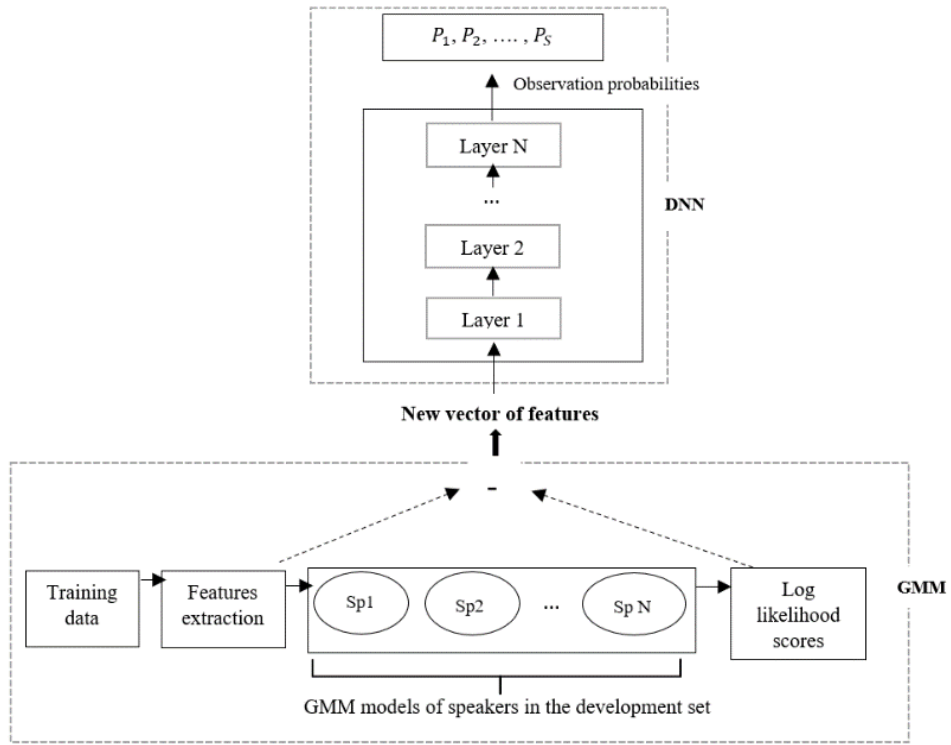

**FIGURE 8.** Architecture of the GMM-DNN model in the training phase

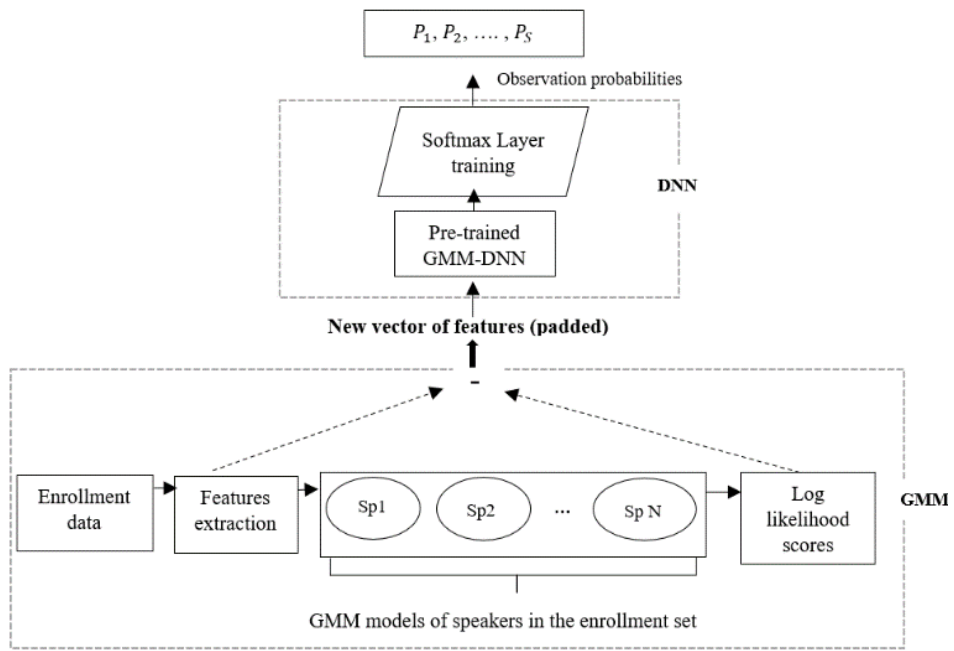

**FIGURE 9.** Architecture of the GMM-DNN model in the enrollment phase

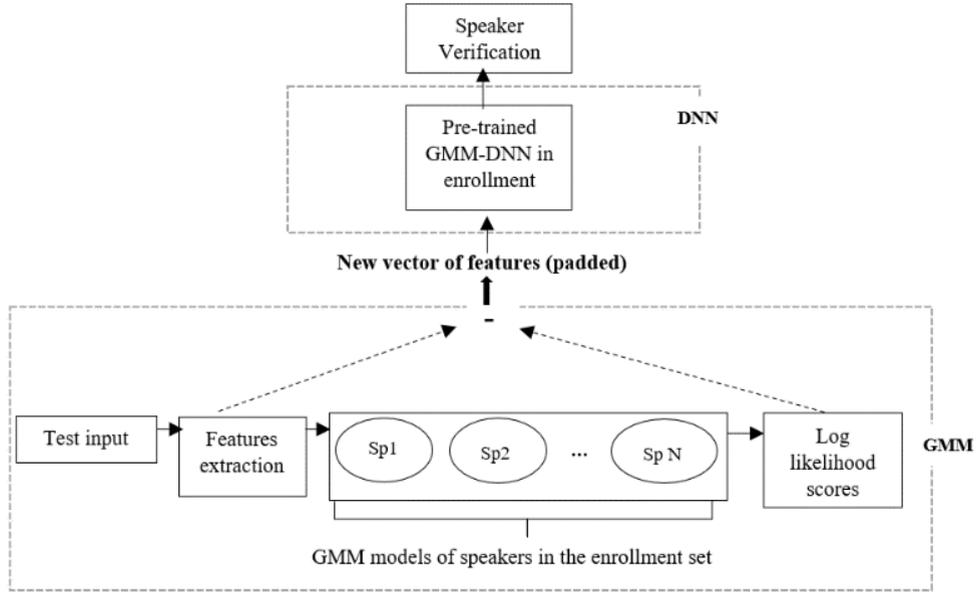

**FIGURE 10. Architecture of the GMM-DNN model in the testing phase**

**4.5. DNN parameters**

In this work, we formed every single DNN hybrid model by stacking multiple dense (fully connected) layers with multiple hidden unit layers between its inputs and outputs. All dense layers used rectified linear unit (ReLU) as activation functions and the he_uniform kernel initializer. The output layer (last layer) is a Softmax layer. All hybrid DNN models were trained using the stochastic gradient descent (SGD) for 150 epochs with the learning rate of categorical cross-entropy set to 0.01 and 0.9 momentum. The batch size was set to 128. Table III provides the hyperparameters used in hybrid modeling of each of DNN-HMM, DNN-GMM, HMM-DNN, and GMM-DNN based on the Emirati database. Experimental trials proved that additional hidden layers did not help in improving performance accuracy.

TABLE III
HYPERPARAMETERS UTILIZED DURING HYBRID MODELING

| Parameters | DNN-HMM | DNN-GMM | HMM-DNN | GMM-DNN |
|---|---|---|---|---|
| Input dimension | 16 | 16 | 24 | 24 |
| #hidden layers | {1,2, output layer} | {1,2, output layer} | {1, output layer} | {1,2, output layer} |
| #neurons per layer | {500, 500, 16} | {500, 500, 16} | {64, 24} | {128, 128, 24} |
| activation | {Relu, Relu, Relu} | {Relu, Relu, Relu} | {Relu, Relu, Relu} | {Relu, Relu, Relu} |
| optimizer | SGD | SGD | SGD | SGD |
| learning rate | 0.01 | 0.01 | 0.01 | 0.01 |
| momentum | 0.9 | 0.9 | 0.9 | 0.9 |

**5. Threshold and Speaker Verification**

In a speaker verification setup, a threshold is used to decide whether an identity claim is genuine or false. Two potential types of errors may emerge: false rejection (FR) and false acceptance (FA). A FR error arises when a genuine speaker's identity claim is rejected, whereas a FA error occurs when the identity claim of an impostor is accepted [9]. EER corresponds to the value where the false rejection rate (FRR) and the false acceptance rate (FAR) are equal. It is usually employed as an evaluation metric in verification tasks and is one of the fundamental parameters in binary decision systems.

In this study, the thresholds are generated using the roc_curve() built-in function in Python. Then, a ROC curve is plotted, which shows the tradeoff of true positive rate (tpr) and false positive rates (fpr) at different threshold settings. Choosing a high threshold acceptance value will cause increased false rejection rates. Likewise, selecting a low threshold value will increase FARs.

The last step in the authentication procedure is to compare the score against the optimal threshold $\theta$ to accept or reject the claimed identity of the speaker, i.e. [7],

$$\begin{cases} if\ score \geq \theta, Accept\ the\ claimed\ identity \\ if\ score < \theta, Reject\ the\ claimed\ identity \end{cases}, \qquad (7)$$

## 6. Experimental Analysis and Discussions

In this work, we developed four different approaches based on hybrid DNN architectures to assess verification performance in emotional and stressful auditory environments.

For the collected Emirati database, 32-dimension of MFCCs, their first and second derivatives (16 static MFCCs and 16 delta MFCCs) are extracted to form the feature vectors. The training stage is comprised of 24 speakers out of 31 speakers (7 males and 17 females) pronouncing 5 out of the 8 sentences which are randomly selected. In this phase, each speaker replicates each sentence 9 times in the neutral state. Hence, the overall number of utterances utilized in the training stage is equal to 1,080 utterances (24 speakers × 5 sentences × 9 replicates × neutral state). The evaluation phase involves each one of the remaining 7 speakers (2 males and 5 females) pronouncing the remaining 3 sentences with 9 replications per sentence under each of the neutral, angry, happy, sad, fear, and disgust emotion. Hence, the test phase encompasses a total of 1,134 utterances (7 speakers × 3 sentences × 9 repetitions × 6 emotions). Five distinct models were generated with different training and test sets for each model. Ultimately, the average value of each of EER and AUC across all models was computed. This ensures that the obtained results are not biased. The percentage EER and AUC scores of each hybrid model using the ESD database are reported in Table IV.

Table IV results clearly show that both HMM-DNN and GMM-DNN demonstrate remarkably low error rates and high AUC values compared to those of DNN-GMM and DNN-HMM models in the neutral as well as in the happy, sad, fear, and disgust acoustic milieus, followed by DNN-GMM and eventually by DNN-HMM. However, GMM-DNN fails to surpass DNN-GMM, with respect to the angry emotion, with EER values equivalent to 8.36% and 4.87%, respectively. When utterances are expressed with anger, GMM-DNN demonstrates an EER value close to that of DNN-HMM, namely, 8.36% and 8.71%, respectively. In addition, the hybrid HMM-DNN model outperforms all hybrid models in terms of error rates and AUC evaluation metrics. On the other hand, the DNN-HMM model demonstrates the worst verification performance. In fact, the former classifier yields an average EER percentage of 0.22% compared to 7.31%, 4.15%, and 1.74% based on DNN-HMM, DNN-GMM, and GMM-DNN, respectively.

Fig. 11 illustrates comparison plots of the ROC curves among DNN-HMM, HMM-DNN, DNN-GMM, and GMM-DNN in different emotional states using the private ESD database. The plots show the tradeoff between false acceptance rate and true acceptance rate obtained using the first created model (five distinct models are generated as mentioned earlier). Plots (a)–(f) demonstrate that the proposed HMM-DNN approach outperforms the other three hybrid models in terms of AUC scores in the neutral and emotional states. The best performance is recorded in neutral-talking conditions, with 0.97, 1.0, 0.99, and 0.99 AUC scores based on DNN-HMM, HMM-DNN, DNN-GMM, and GMM-DNN, respectively. Plot 10-b demonstrates that anger provides the worst ROC curve, with the least AUC score achieved by GMM-DNN: 0.92 compared to 0.96, 1, and 0.98 for DNN-HMM, HMM-DNN, and DNN-GMM, respectively.

TABLE IV
SPEAKER VERIFICATION PERFORMANCE IN EMOTIONAL TALKING ENVIRONMENTS USING THE ESD DATABASE BASED ON PERCENTAGE EER AND AUC VALUES

| Emotion | DNN-HMM | | HMM-DNN | | DNN-GMM | | GMM-DNN | |
|---|---|---|---|---|---|---|---|---|
| | EER | AUC | EER | AUC | EER | AUC | EER | AUC |
| Neutral | 6.55 | 0.97 | **0.01** | **1** | 3.74 | 0.99 | 0.30 | 0.99 |
| Angry | 8.71 | 0.96 | **0.73** | **0.99** | 4.87 | 0.98 | 8.36 | 0.95 |
| Happy | 7.25 | 0.97 | **0.04** | **1** | 4.18 | 0.98 | 0.70 | 0.99 |
| Sad | 7.08 | 0.97 | **0.00** | **1** | 3.94 | 0.97 | 0.01 | 1 |
| Fear | 7.12 | 0.97 | **0.00** | **1** | 3.95 | 0.98 | 0.06 | 1 |
| Disgust | 7.19 | 0.97 | **0.57** | **0.99** | 4.26 | 0.98 | 0.99 | 0.99 |
| **Average** | **7.31** | **0.97** | **0.22** | **0.99** | **4.15** | **0.98** | **1.74** | **0.99** |

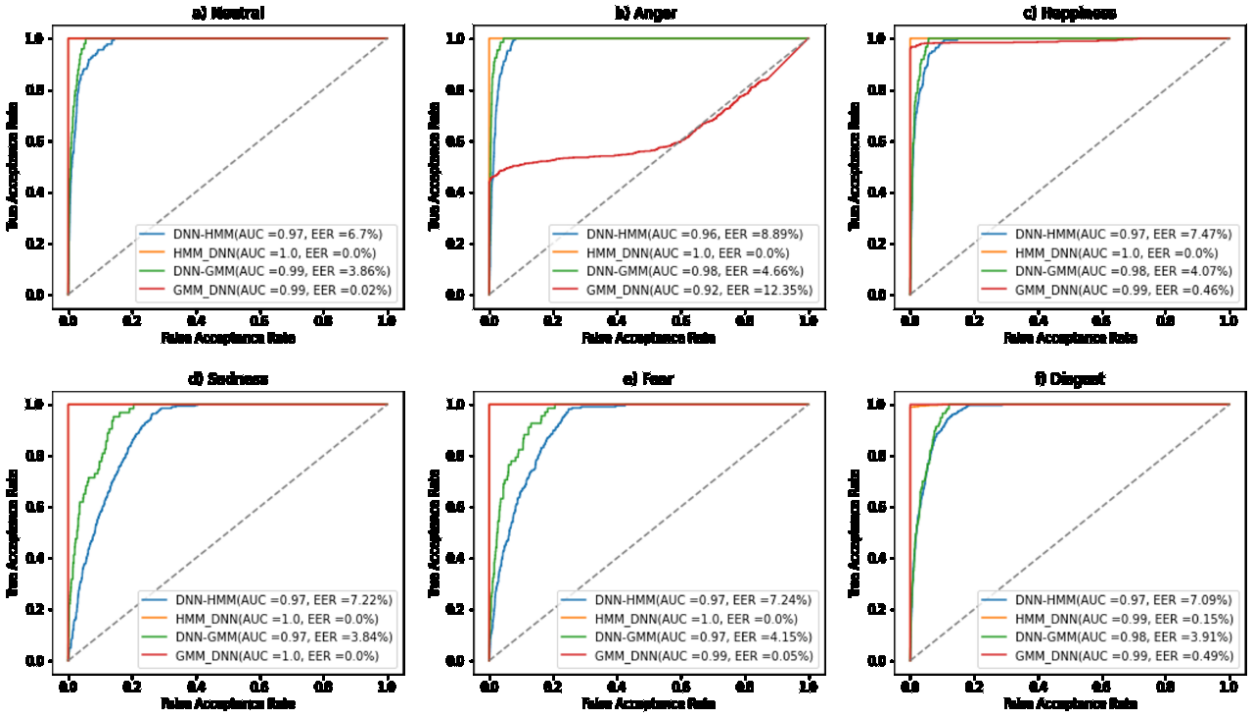

**FIGURE 11.** ROC plots for each of DNN-HMM, HMM-DNN, DNN-GMM, and GMM-DNN for the different emotional states for model #1 using the ESD database

TABLE V
PERCENTAGE EER AND AUC USING GMM, HMM, AND DNN CLASSIFIERS USING THE ESD DATABASE

|  | GMM | | HMM | | DNN | |
|---|---|---|---|---|---|---|
|  | **EER** | **AUC** | **EER** | **AUC** | **EER** | **AUC** |
| Neutral | 5.03 | 0.97 | 2.33 | 0.98 | **1.22** | **0.99** |
| Angry | 22.65 | 0.85 | 16.67 | 0.91 | **11.75** | **0.94** |
| Happy | 12.50 | 0.94 | 7.74 | 0.96 | **6.14** | **0.97** |
| Sad | 8.99 | 0.97 | 5.19 | 0.98 | **4.09** | **0.98** |
| Fear | 13.97 | 0.94 | 9.10 | 0.95 | **7.15** | **0.96** |
| Disgust | 6.22 | 0.96 | 7.33 | 0.96 | **5.30** | **0.97** |
| **Average** | **11.56** | **0.94** | **8.06** | **0.96** | **5.94** | **0.97** |

On the basis of the results shown in Table IV and Table V, it is clear that the hybrid approaches using the private Emirati database outperform GMM, HMM, and DNN solo classifiers in terms of EER and AUC values. In fact, the implemented HMM-DNN approach achieves a remarkable decrease in the error rate percentage equivalent to 98.09%, 97.27%, and 96.29% for GMM, HMM, and DNN, respectively.

We carried out the following six experiments in an autonomous fashion to validate the results achieved using the Emirati dataset:

**6.1. Experiment #1 SUSAS database**

The SUSAS dataset is used in this experiment to assess the verification performance of the hybrid models. The training phase consists of 6 out of 9 male speakers (3 speakers with a Boston regional accent and 3 other speakers with a General American accent). During this phase, each speaker utters 30 out of 35 diverse sentences with 2 repetitions per sentence under the neutral state. Therefore, the total number of utterances is 360 (6 speakers × 30 sentences × 2 repetitions/sentence × neutral state) are allocated in the training stage. The remaining 3 speakers, with a New York regional accent, are designated for the test phase. During this phase, each test speaker articulates the remaining 5 words under each of the 6 stressful conditions, which are neutral, angry, slow, soft, Lombard, and fast. Overall, 180 speech samples (3 speakers × 5 sentences × 2 repetitions/sentence × 6 stressful states) are used in the testing stage. The input speech signals are converted into MFCCs and delta–deltas of 40 dimensions.

On the basis of the results shown in Table VI, we deduce that the HMM-DNN outperformed all other hybrid models in terms of EER at the neutral as well as the different stressful speaking styles. The highest error rates were registered at both angry and the Lombard speaking styles based on the four hybrid classifiers. For angry, the EERs were equivalent to 9.67%, 31.06%, 17.83%, and 34.0% based on HMM-DNN, DNN-HMM, DNN-GMM, and GMM-DNN, respectively. For the Lombard speaking style, the EERs were 10.58%, 31.06%, 11.92%, and 21.0% based on HMM-DNN, DNN-HMM, DNN-GMM, and GMM-DNN, respectively. Conversely, DNN-HMM displayed the poorest average verification performance, with an EER value of 19.98% compared to 3.67%, 9.55%, and 10.02% for HMM-DNN, DNN-GMM, and GMM-DNN, respectively. These results match those derived using the emotional Emirati database.

Fig. 12 depicts the graphical plots of the ROC curves for verification performance obtained using the SUSAS database at different threshold settings based on DNN-HMM, HMM-DNN, DNN-GMM, and GMM-DNN. We deduce that in the neutral state, the calculated AUC value of the HMM-DNN model was higher than that of each hybrid model: 1 compared to 0.94, 0.98, and 0.99 for DNN-HMM, DNN-GMM, and GMM-DNN, respectively. Similarly, HMM-DNN beats all hybrid approaches at various stressful speaking styles. As per the plots (b) and (f), the verification performance is the poorest under both angry and Lombard talking conditions. In fact, all hybrid models attained the lowest AUC values in these two conditions in comparison with neutral, fast, slow, and soft speaking styles. For the angry condition, HMM-DNN recorded an AUC value equivalent to 0.93 against 0.74, 0.80, and 0.68 based on DNN-HMM, DNN-GMM, and GMM-DNN, respectively. For the Lombard, the AUC value is 0.95, 0.64, 0.71, and 0.86 based on HMM-DNN, DNN-HMM, DNN-GMM, and GMM-DNN, respectively. In addition, we conclude that DNN-HMM and DNN-GMM achieved the lowest AUC values across all emotional states with ROC curves being close to the diagonal indicating the poor performance of both DNN-HMM and DNN-GMM compared to HMM-DNN and GMM-DNN.

TABLE VI
SPEAKER VERIFICATION PERFORMANCE IN STRESSFUL TALKING ENVIRONMENTS USING SUSAS DATABASE BASED ON PERCENTAGE EER AND AUC VALUES

| Emotion | DNN-HMM | | HMM-DNN | | DNN-GMM | | GMM-DNN | |
|---|---|---|---|---|---|---|---|---|
| | EER | AUC | EER | AUC | EER | AUC | EER | AUC |
| Neutral | 11.44 | 0.94 | **0.08** | **1** | 6.22 | 0.98 | 2.33 | 0.99 |
| Angry | 31.06 | 0.74 | **9.67** | **0.93** | 17.83 | 0.80 | 34.0 | 0.68 |
| Fast | 13.2 | 0.94 | **0.33** | **0.99** | 5.77 | 0.97 | 1.75 | 0.99 |
| Slow | 16.16 | 0.91 | **0.33** | **1** | 7.92 | 0.96 | 0.67 | 0.99 |
| Soft | 16.96 | 0.90 | **1.08** | **0.99** | 7.65 | 0.92 | 0.42 | 0.99 |
| Lombard | 31.14 | 0.64 | **10.58** | **0.95** | 11.92 | 0.71 | 21.0 | 0.86 |
| **Average** | **19.98** | **0.84** | **3.67** | **0.97** | **9.55** | **0.89** | **10.02** | **0.91** |

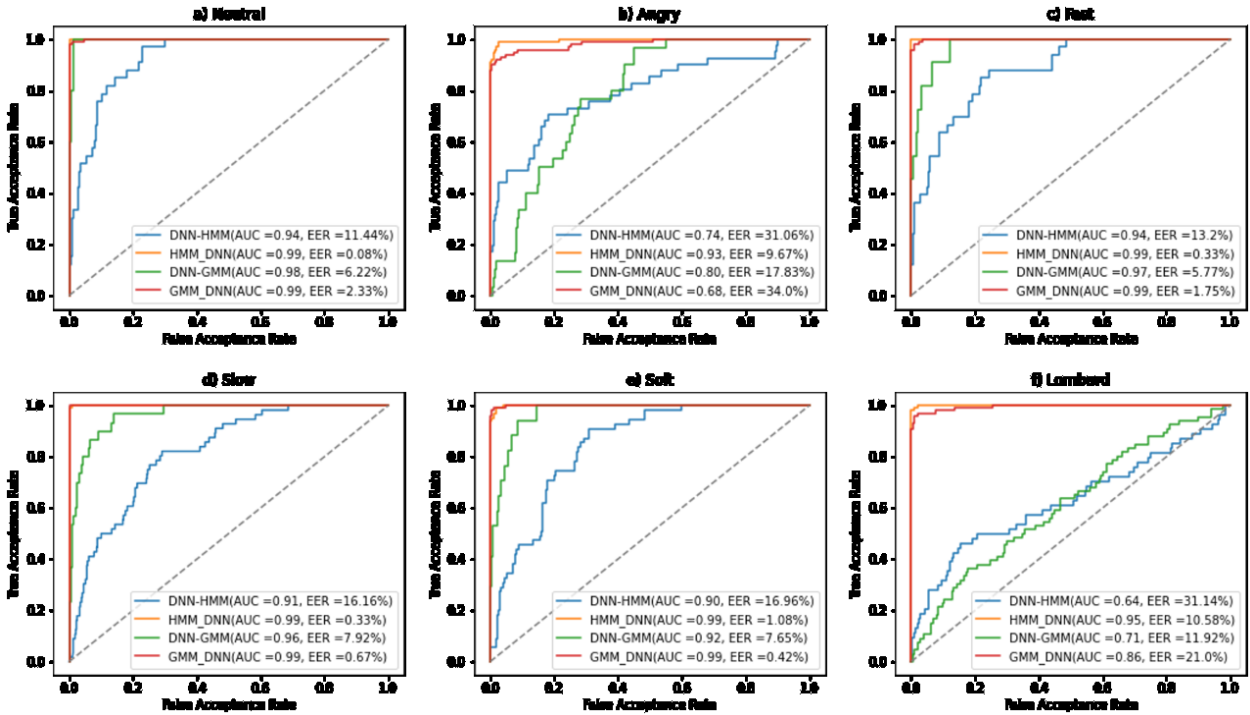

**FIGURE 12.** ROC plots for each of DNN-HMM, HMM-DNN, DNN-GMM, and GMM-DNN for the different stressful states using SUSAS database

The EER and AUC values shown in Table VI and Table VII clearly demonstrate that the hybrid classifiers demonstrate superior average verification performance compared to those achieved by GMM alone, HMM alone, and DNN alone classifiers. The decreased average error rates in the proposed HMM-DNN were 94.79%, 93.45%, and 91.18% for GMM, HMM, and DNN, respectively.

TABLE VII
PERCENTAGE EER AND AUC USING GMM, HMM, AND DNN CLASSIFIERS USING SUSAS DATABASE

|  | GMM | | HMM | | DNN | |
| --- | --- | --- | --- | --- | --- | --- |
|  | EER | AUC | EER | AUC | EER | AUC |
| Neutral | 63.33 | 0.30 | 53.33 | 0.44 | **33.66** | **0.69** |
| Angry | 63.33 | 0.24 | 53.33 | 0.32 | **45.28** | **0.56** |
| Fast | 63.33 | 0.30 | 50.00 | 0.51 | **34.94** | **0.67** |
| Slow | 80.00 | 0.14 | 60.00 | 0.32 | **39.38** | **0.62** |
| Soft | 93.33 | 0.33 | 70.00 | 0.24 | **42.32** | **0.60** |
| Lombard | 60.00 | 0.31 | 50.00 | 0.38 | **50.21** | **0.51** |
| **Average** | **70.56** | **0.21** | **56.11** | **0.37** | **41.65** | **0.61** |

**6.2. Experiment #2 RAVDESS database**

The RAVDESS database is used in this experiment to evaluate the performance of the four DNN-based hybrid models. The training stage is made up of the first 20 actors and actresses (10 males and 10 females) out of the 24 actors and actresses expressing the first sentence, out of 2 sentences, with a repetition of 2 times per sentence in the neutral emotion. Therefore, a total of 40 speech samples (20 speakers × 1 sentence × 2 trials/sentence × neutral state) are gathered for training. The remaining 4 actors and actresses (2 males and 2 females) are designated for evaluation. The evaluation stage involves the second utterance expressed in each of angry, sad, happy, fear, and disgust emotional classes in addition to the neutral state. Thus, the size of the dataset in this phase is equivalent to 88 utterances [(4 speakers × 1 utterance × 2 trials/utterance × neutral state) + (4 speakers × 1 utterance × 4 trials/utterance × 5 emotional states)].

Based on the average EER and AUC results shown in Table VIII, it is obvious that HMM-DNN performed the best among hybrid models followed by DNN-GMM, DNN-HMM, and GMM-DNN. The average EER percentages recorded were 17.70% based on HMM-DNN compared to 23.26%, 20.83%, and 23.95% based on DNN-HMM, DNN-GMM, and GMM-DNN, respectively. However, the EER and AUC scores of each emotional category alone demonstrate that HMM-DNN failed to beat other hybrid models in the neutral, angry, happy, and disgust. The obtained results are slightly biased since the RAVDESS has a very small training dataset (one utterance out of a total of two utterances). For instance, the neutral state attains an EER of 0% and an AUC of 1 based on each of DNN-HMM and DNN-GMM which seems to be ideal for real-world scenario.

Fig. 13 shows the plots of the ROC curves that compare the verification performance based on each hybrid architecture at different classification thresholds using the RAVDESS database. The plots (a)–(f) clearly show that HMM-DNN reached better verification performance with respect to sad and fear emotions than all other emotions. For the sad emotion, the AUC values are equivalent to 0.93, 0.91, 0.70, and 0.93 based on HMM-DNN, DNN-HMM, DNN-GMM, and GMM-DNN, respectively. For the fear emotion, the AUC recorded 0.97, 0.64, 0.75, and 0.85 based on HMM-DNN, DNN-HMM, DNN-GMM, and GMM-DNN, respectively. However, the DNN-HMM gave the best ROC curves in the angry and happy emotions. When considering the average AUC scores achieved, we conclude that HMM-DNN surpasses all hybrid models; 0.80 beats 0.77, 0.79, and 0.76 based on DNN-HMM, DNN-GMM, and GMM-DNN, respectively.

TABLE VIII

SPEAKER VERIFICATION PERFORMANCE IN EMOTIONAL TALKING ENVIRONMENTS USING RAVDESS DATABASE BASED ON PERCENTAGE EER AND AUC VALUES

| Emotion | DNN-HMM | | HMM-DNN | | DNN-GMM | | GMM-DNN | |
| --- | --- | --- | --- | --- | --- | --- | --- | --- |
| | EER | AUC | EER | AUC | EER | AUC | EER | AUC |
| Neutral | 0.00 | 1 | **4.17** | **0.95** | 0.00 | 1 | 4.17 | 0.95 |
| Angry | 33.33 | 0.70 | **45.83** | **0.45** | 33.33 | 0.66 | 50.00 | 0.50 |
| Sad | 14.58 | 0.91 | **4.17** | **0.93** | 29.17 | 0.70 | 12.50 | 0.93 |
| Happy | 14.58 | 0.89 | **22.92** | **0.83** | 20.83 | 0.79 | 27.08 | 0.77 |
| Fear | 37.50 | 0.64 | **2.08** | **0.97** | 25.00 | 0.75 | 14.58 | 0.85 |
| Disgust | 39.58 | 0.50 | **27.08** | **0.70** | 16.67 | 0.86 | 35.42 | 0.66 |
| **Average** | **23.26** | **0.77** | **17.70** | **0.80** | **20.83** | **0.79** | **23.95** | **0.76** |

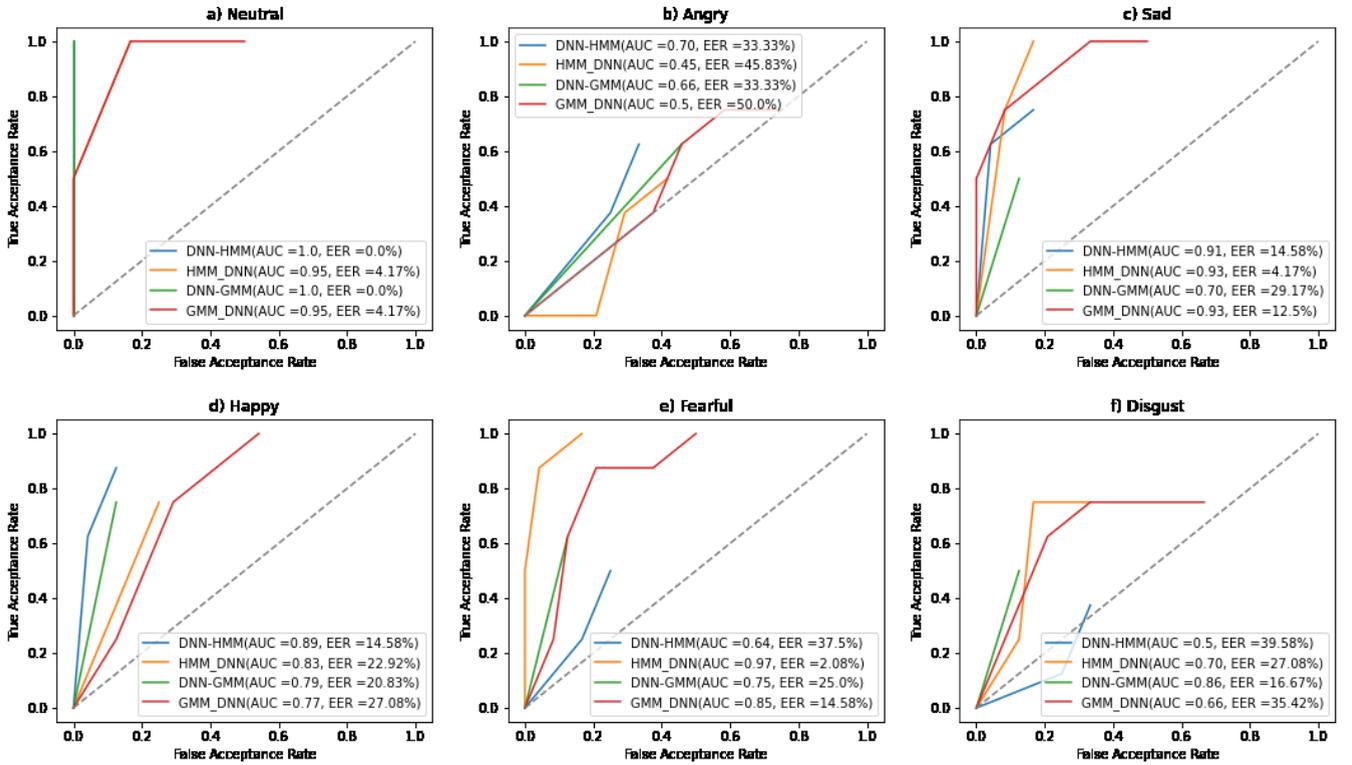

*FIGURE 13. ROC plots for each of DNN-HMM, HMM-DNN, DNN-GMM, and GMM-DNN for the different emotional states using RAVDESS database*

From both Table VIII and Table IX, it is evident that HMM-DNN outperforms all solo classifiers. The proposed HMM-DNN demonstrated improved EERs equal to 50.02%, 50.02%, and 31.10% over GMM, HMM, and DNN, respectively.

TABLE IX
PERCENTAGE EER AND AUC USING GMM, HMM, AND DNN CLASSIFIERS USING RAVDESS DATABASE

| Emotion | GMM | | HMM | | DNN | |
|---|---|---|---|---|---|---|
| | **EER** | **AUC** | **EER** | **AUC** | **EER** | **AUC** |
| Neutral | 0.00 | 0.87 | 0.00 | 0.87 | **0.00** | **0.95** |
| Angry | 50.00 | 0.39 | 50.00 | 0.39 | **39.58** | **0.64** |
| Happy | 37.50 | 0.51 | 37.50 | 0.51 | **37.50** | **0.71** |
| Sad | 25.00 | 0.65 | 25.00 | 0.65 | **14.58** | **0.84** |
| Fear | 50.00 | 0.45 | 50.00 | 0.45 | **25.00** | **0.72** |
| Disgust | 50.00 | 0.46 | 50.00 | 0.46 | **37.50** | **0.69** |
| **Average** | **35.42** | **0.55** | **35.42** | **0.55** | **25.69** | **0.76** |

**6.3. Experiment #3 Time complexity performance study**

Table X lists the average calculated training and testing times, measured in seconds, of the hybrid classifiers as well as of HMM, GMM, DNN, SVM, and MLP classifiers in each of the three databases. The training/testing time measures the elapsed time between the beginning and the ending of each operation. In this work, the Python built-in time method *time.time()* is utilized.

In light of hybrid models, we observed that DNN-GMM model achieved the fastest training time compared to all other hybrid models in the three datasets. Using the ESD database, DNN-GMM reached 186.683 s compared to DNN-HMM, HMM-DNN, and GMM-DNN, which reached 189.187, 325.825, and 190.321 s, respectively. Using the RAVDESS database, the recorded training times were 3.506, 4.615, 4.489, and 4.476 s based on DNN-GMM, DNN-HMM, HMM-DNN, and GMM-DNN, respectively. Using the SUSAS dataset, the training times were equivalent to 120.314, 223.694, 192.734, and 182.085 s based on DNN-GMM, DNN-HMM, HMM-DNN, and GMM-DNN, respectively.

Using the ESD database, the DNN-GMM required the least time, 293.74 s, compared to DNN-HMM, HMM-DNN, and GMM-DNN, which required 783.861, 637.645, and 304.155 s. Likewise, using the RAVDESS dataset, DNN-GMM required the least testing times compared to DNN-GMM, DNN-HMM, HMM-DNN, and GMM-DNN, which required 1.794, 1.876, 1.849, and 1.944 s, respectively. Similarly, using the SUSAS database, DNN-GMM required the least testing time with 60.602 s compared to DNN-GMM, DNN-HMM, HMM-DNN, and GMM-DNN, which required 85.740, 170.101, and 71.867 s, respectively.

On the basis of the Table X results, all examined hybrid models have higher computational costs associated with training and testing times than those of HMM, GMM, and DNN solo classifiers. This is due to the cascaded nature of the hybrid models that necessitates the training/testing of the first classifier, followed by the training/testing of the second classifier.

The computational times of the examined hybrid models are compared to those of the following classical classifiers: SVM and MLP. We deduce from Table X results that both SVM and MLP demonstrate the largest time complexity with regard to training and testing times compared to hybrid models across the three databases.

Upon conducting some experiments with respect to the training and testing times, our findings demonstrate that when the size of the dataset increases, the corresponding training and testing times increase, but not in a proportional manner.

TABLE X
AVERAGE TRAINING AND TESTING TIME (IN SECONDS) FOR THE ESD, RAVDESS, AND SUSAS DATABASES

| Classifier | ESD | | RAVDESS | | SUSAS | |
|---|---|---|---|---|---|---|
| | **Training time** | **Testing time** | **Training time** | **Testing time** | **Training time** | **Testing time** |
| DNN-HMM | 189.187 | 783.861 | 4.615 | 1.876 | 223.694 | 85.740 |
| HMM-DNN | 325.825 | 637.645 | 4.489 | 1.849 | 192.734 | 170.101 |
| DNN-GMM | 186.683 | 293.740 | 3.506 | 1.794 | 120.314 | 60.602 |
| GMM-DNN | 190.321 | 304.155 | 4.476 | 1.944 | 182.085 | 71.867 |
| HMM | 100.858 | 70.765 | 0.878 | 1.789 | 53.125 | 37.550 |
| GMM | 24.930 | 1.086 | 0.247 | 10.934 | 6.157 | 36.100 |
| DNN | 150.243 | 80.032 | 3.660 | 1.792 | 110.252 | 50.453 |
| MLP | 354.714 | 603.231 | 5.538 | 138.731 | 317.450 | 190.97 |
| SVM | 512.074 | 1805.83 | 5.200 | 142.97 | 300.918 | 200.81 |

**6.4. Experiment #4 Wilcoxon test**

To validate our results and verify whether the obtained error rates are factual or merely caused by statistical fluxes, we carried out statistical significance tests between the HMM-DNN (the winning model) and the rest of the DNN-based hybrid models. On the basis of the Kolmogorov–Smirnov test, we deduced that the error rates are not normally distributed. To this end [39], we utilized the Wilcoxon test, which is nonparametric.

Table I provides the p-values achieved for the Emirati database. We observe that every calculated p-value among HMM-DNN and GMM-DNN and DNN-GMM is less than α = 0.05 in both neutral and emotional talking environments. Therefore, HMM-DNN statistically differs from both DNN-GMM and GMM-DNN. In addition, we remark that there is a statistical difference between HMM-DNN and DNN-HMM in neutral, sad, happy, fear, and disgust emotional environments. However, HMM-DNN does not differ from DNN-HMM in the angry emotion.

TABLE XI
P-VALUES BETWEEN HMM-DNN AND OTHER MODELS USING WILCOXON TEST FOR THE EMIRATI DATABASE (α = 0.05)

| **Emotion** | **DNN-HMM** | **GMM-DNN** | **DNN-GMM** | |
|---|---|---|---|---|
| Neutral | **0.000** | **0.000** | **0.000** | |
| Angry | 0.087 | **0.000** | **0.024** | |
| Sad | **0.009** | **0.000** | **0.000** | **HMM-DNN** |
| Happy | **0.000** | **0.000** | **0.000** | |
| Fear | **0.005** | **0.000** | **0.000** | |
| Disgust | **0.000** | **0.000** | **0.000** | |

Table XII presents the calculated p-values for the SUSAS database using the Wilcoxon test. For alpha (α) = 0.05, we notice that 1) HMM-DNN differs from all hybrid models in neutral speech, 2) HMM-DNN and DNN-HMM significantly differ, except for the angry speaking style, 3) HMM-DNN and GMM-DNN statistically differ in all stressful speaking styles, and 4) HMM-DNN differs from DNN-GMM, except for the slow speaking style.

TABLE XII
P-VALUE BETWEEN HMM-DNN AND OTHER MODELS USING WILCOXON TEST FOR THE SUSAS DATABASE (α = 0.05)

| Emotion | DNN-HMM | GMM-DNN | DNN-GMM | |
|---|---|---|---|---|
| Neutral | **0.005** | **0.000** | **0.000** | |
| Angry | 0.350 | **0.000** | **0.000** | |
| Fast | **0.000** | **0.000** | **0.000** | **HMM-DNN** |
| Slow | **0.000** | **0.000** | 0.353 | |
| Soft | **0.000** | **0.000** | **0.000** | |
| Lombard | **0.000** | **0.000** | **0.000** | |

Table XIII lists the p-values obtained for the RAVDESS database. Results show that HMM-DNN and DNN-GMM significantly differ for α = 0.1, except for the sad and fear emotions. However, HMM-DNN does not differ from GMM-DNN in all emotional states. Likewise, the winning hybrid model (HMM-DNN) does not statistically differ from DNN-HMM in the neutral, sad, happy, and disgust emotional states.

TABLE XIII
P-VALUE BETWEEN HMM-DNN AND OTHER MODELS USING WILCOXON TEST FOR THE RAVDESS DATABASE (α = 0.1)

| Emotion | DNN-HMM | GMM-DNN | DNN-GMM | |
|---|---|---|---|---|
| Neutral | 0.157 | 0.674 | **0.002** | |
| Angry | **0.014** | 0.366 | **0.000** | |
| Sad | 0.690 | 0.140 | 0.125 | **HMM-DNN** |
| Happy | 0.484 | 0.945 | **0.002** | |
| Fear | **0.009** | 0.379 | 0.207 | |
| Disgust | 0.116 | 0.277 | **0.004** | |

## 6.5. Experiment #5 Comparison with prior work

In this experiment, the average speaker verification performance achieved using our proposed HMM-DNN hybrid classifier was compared with the verification results reported by previous research using different types of classifiers such as HMM1, HMM2, and HMM3 as in [7]; CHMM3 and CSPHMM employed by Shahin and Nassif in [4]; and (EVA)-based i-vector implemented by Prasetio et al. in [17]. The comparison was restricted to studies that utilized the Emirati database, the SUSAS database, or the RAVDESS dataset in their experiments in the neutral, emotional, or stressful talking conditions.

Table XIV demonstrates the average EER percentage values reported by previous studies as well as the error rate percentage decreases that our newly proposed HMM-DNN achieved. On the basis of the results of this table, it is evident that HMM-DNN has an advantage over HMM1, HMM2, HMM3, CHMM3, CSPHMM, and EVA-based i-vector. In fact, the proposed HMM-DNN results in a significant reduction in the EER values using the Emirati database in neutral and emotional talking environments. The percentage decrease was equivalent to 99.91%, 99.89%, 99.79%, 99.24%, and 98.98% for HMM1, HMM2, HMM3, CHMM3, and CSPHMM, respectively. Similarly, using the SUSAS database, HMM-DNN led to a reduction in error rate percentage in the stressful talking condition equivalent to 18.62%, 16.01%, and 10.04% based on CSS, EDS, and MDS, respectively.

The percentage decrease in EER rate is calculated using the following equation:

$$\text{Percentage decrease in EER} = \frac{\text{EER of previous work} - \text{EER of proposed model}}{\text{EER of previous work}} \times 100\% \qquad (8)$$

TABLE XIV
PERCENTAGE EER VALUES ACHIEVED BY PREVIOUS STUDIES AS WELL AS THE PERCENTAGE DECREASE IN ERROR RATES REPORTED USING THE PROPOSED HMM-DNN

| Reference | Talking condition | Database | Classifier | EER of previous work | EER of proposed model | Percentage decrease in error rate |
|---|---|---|---|---|---|---|
| [7] | Neutral only | Emirati | HMM1 | 11.5% | 0.01% | 99.91% |
|  |  |  | HMM2 | 9.6% |  | 99.89% |
|  |  |  | HMM3 | 4.9% |  | 99.79% |
| [4] | Emotional | Emirati | CHMM3 | 29% | 0.22% | 99.24% |
|  |  |  | CSPHMM | 21.75% |  | 98.98% |
| [17] | Stressful | SUSAS | EVA-based i-vector | CSS: 4.51% | 3.67% | 18.62% |
|  |  |  |  | EDS: 4.37% |  | 16.01% |
|  |  |  |  | MDS: 4.08% |  | 10.04% |

## 6.6. Experiment #6 Comparison between emotional and stressful talking environments

In this experiment, we compared the average verification performance between each emotional and stressful environments based on the four DNN-based hybrid models. On the basis of the Table II and Table VI results, the average emotional verification performance attained using the private emotional Emirati database based on each hybrid model was noticeably greater than the average stressful verification performance achieved using the stressful SUSAS database. On the basis of the ESD dataset, the percentage decreases in error rates for HMM-DNN, DNN-HMM, DNN-GMM, and GMM-DNN were 94.00%, 63.41%, 56.54%, and 82.63%, respectively. However, according to the reported EER and AUC scores of Table VI and Table IIIIII, we deduce that the emotional performance achieved using the RAVDESS database is inferior to that achieved by the stressful verification performance. A percentage increase in error rates reached 79.26%, 14.10%, 54.15%, and 58.16% based on HMM-DNN, DNN-HMM, DNN-GMM, and GMM-DNN, respectively, using the RAVDESS database. We conclude from the above discussions that EER and AUC scores are database-dependent, which means that they depend on the database that is being used.

On the basis of the training and testing times of the hybrid models reported in Table X, DNN-GMM achieved the lowest computational time in terms of both training and testing phases when compared to those of other hybrid models in emotional and stressful talking environments.

For performance comparisons, some of the advantages and limitations of the traditional as well as the hybrid classifiers that have been utilized in this work are presented in Table XV. The classifiers are the following: SVM, MLP, HMM, GMM, DNN-HMM, HMM-DNN, DNN-GMM, and GMM-DNN.

TABLE XV
ADVANTAGES AND DISADVANTAGES OF DIFFERENT CLASSIFIER TECHNIQUES [40],[41],[42],[43],[44],[45],[46]

| Type of methods | Classifier | Advantages | Disadvantages |
|---|---|---|---|
| **Classical methods** | **SVM** | ▪ Simple operations [40], [43].<br>▪ Performs well with complex nonlinear data points [41]. | ▪ Not suitable for large databases [40].<br>▪ The main problem is the selection of the appropriate kernel function as it is database-dependent [41]. |
| | **MLP** | ▪ Easy to design and to implement, requires few parameters [41].<br>▪ More effective in modeling nonlinear mappings [41].<br>▪ Flexible in changing environments [44]. | ▪ High computational cost [45].<br>▪ Long training time [45]. |
| | **HMM** | ▪ Strong statistical foundation [41].<br>▪ Efficient performance [40].<br>▪ Able to capture speaker temporal information [41].<br>▪ Provides better classification accuracies than all other classical classifiers. | ▪ Problem with features selection process [41].<br>▪ Computationally more complex, require more storage space [40]. |
| | **GMM** | ▪ Very efficient in modeling multimodal distributions [41], [43].<br>▪ Requires less training and testing data [40], [41], [43].<br>▪ Requires less training and testing times compared to other classical methods. | ▪ Unable to model temporal structure of the training data as all equations relevant to training and testing are founded on the assumption that all vectors are independent [41]. |
| **Hybrid DNN methods** | **DNN-HMM** | ▪ Better average verification performance than DNN, GMM, and HMM. | ▪ Least accurate average verification performance compared to DNN-GMM, HMM-DNN, and GMM-DNN hybrid models in terms of EER and AUC.<br>▪ Higher computational cost than DNN, GMM, and HMM [46].<br>▪ Complex architecture compared to DNN, GMM, and HMM. |
| | **DNN-GMM** | ▪ Better average verification performance than GMM-DNN in SUSAS and RAVDESS databases.<br>▪ Has the smallest computational complexity of both training and testing phases among hybrid models. | ▪ Less average verification performance than GMM-DNN in ESD database.<br>▪ Complex architecture compared to DNN, GMM, and HMM.<br>▪ Longer training and testing times than DNN, GMM and HMM. |
| | **HMM-DNN** | ▪ Best average accuracy performance compared to all hybrid models in terms of EER and AUC.<br>▪ Outperforms in both emotional and stressful environments. | ▪ Slowest computational cost for both training and testing times in the emotional as well as the stressful talking conditions.<br>▪ Complex architecture than DNN-HMM and DNN-GMM.<br>▪ Complex architecture compared to DNN, GMM, and HMM. |
| | **GMM-DNN** | ▪ Better average verification performance than DNN-HMM across all datasets.<br>▪ Less training and testing times compared to that of HMM-DNN and DNN-HMM. | ▪ Longer training time than DNN-GMM.<br>▪ Complex architecture than DNN-HMM and DNN-GMM.<br>▪ Complex architecture compared to DNN, GMM, and HMM. |

# 7. Conclusion

In this paper, we have investigated the speaker verification performance, in a text-independent setting, within four hybrid DNN-based architectures in each of the emotional and stressful acoustic atmospheres, as a first initiative of its own kind. The hybrid models considered are: the novel HMM-DNN, DNN-HMM, GMM-DNN, and DNN-GMM. Six distinct experiments have been conducted in order to analyze and evaluate the verification performance of the hybrid DNN-based models using three distinct speech databases. This work demonstrates that the novel proposed, designed and implemented HMM-DNN is the best among hybrid models in achieving low error rates and high AUC scores in both emotional and stressful speaker verification systems. In addition, the proposed HMM-DNN demonstrates better verification performance than GMM, HMM, and DNN alone across all datasets. Moreover, DNN-GMM were found to attain the lowest computational time in terms of both training and testing phases when compared to other hybrid models at the emotional as well as the stressful talking conditions. The HMM-DNN model required the greatest amount of time in the training phase. When compared to the traditional SVM and MLP, DNN-GMM achieves significantly low testing time in both emotional and stressful talking conditions. Furthermore, the proposed HMM-DNN is superior to previous research in emotional and stressful taking environments. Moreover, findings proved that EER and AUC scores are database-dependent when comparing stressful and emotional auditory environments.

There are some limitations in this work. First, the proposed HMM-DNN is deemed a computationally expensive model when it comes to the calculated training and testing times compared to HMM, GMM and DNN. In addition, HMM-DNN fails to achieve lower computational cost than all other hybrid models. Also, HMM-DNN falls short as it fails to be statistically different from DNN-GMM at all emotional states and from DNN-HMM at neutral, happy, sad, and disgust using the RAVDESS dataset. Besides, HMM-DNN is statistically not different from DNN-HMM in the angry emotion for Emirati and SUSAS speech corpora.

As future endeavors, our research will be focused on implementing novel HMM-DNN approaches with reduced computational complexity. Furthermore, we aim to scrutinize the behavior of the proposed HMM-DNN model with respect to statistical significance test in small size datasets as well as large size databases. In addition, we aim to develop modified Capsule Network models in the context of speaker verification in abnormal talking environments.

**Compliance with Ethical Standards**

The authors would like to convey their thanks and appreciation to the "University of Sharjah" for supporting the work through the research group – Machine Learning and Arabic Language Processing

**Conflict of Interest:** The authors declare that they have no conflict of interest.

**Informed consent:** This study does not involve any experiments on animals.

**Statement of Ethics:** The authors have authorization from the "University of Sharjah to gather speech database from UAE nationals based on the competitive research project entitled Emirati-Accented Speaker and Emotion Recognition Based on Deep Neural Network, No. 19020403139."